\documentclass[%
notitlepage,%
onecolumn,%
oneside,%
floats,%
aps,%
prd,%
nobibnotes,%
nofootinbib,%
amsmath,%
amssymb,%
amsfonts,%
amscd,%
superscriptaddress,%
eqsecnum%
]{revtex4-1}

\usepackage[utf8]{inputenc}
\usepackage{graphicx, array, dcolumn}
\usepackage[paperwidth=210mm, paperheight=297mm, centering,
hmargin=2.0cm, vmargin=2.4cm]{geometry}

\begin{document}

\title{$f_1 (1285) \to e^+e^-$ decay and direct $f_1$ production in $e^+e^-$ collisions}

\author{A.S.\,Rudenko}
\email{a.s.rudenko@inp.nsk.su}
\affiliation{Budker Institute of Nuclear Physics, Novosibirsk 630090, Russia}
\affiliation{Novosibirsk State University, Novosibirsk 630090, Russia}

\begin{abstract}

The width of the $f_1 (1285) \to e^+e^-$ decay is calculated in the
vector meson dominance model. The result depends on the relative phase between
two coupling constants describing $f_1 \to \rho^0\gamma$ decay. 
The width $\Gamma (f_1 \to e^+e^-)$ is estimated to be $\simeq 0.07 \textendash 0.19$ eV.
Direct $f_1$ production in $e^+e^-$ collisions is discussed, and
the $e^+e^- \to f_1\to a_0 \pi \to \eta \pi \pi$ cross section is calculated.
Charge asymmetry in the $e^+e^- \to \eta \pi^+ \pi^-$ reaction due to
interference between $e^+e^- \to f_1$ and $e^+e^- \to \eta \rho^0$ amplitudes
is studied.
\end{abstract}

\maketitle

\section{Introduction}

High-luminosity electron-positron colliders are powerful tools for measuring
electronic widths of hadronic resonances with positive charge parity, $C=+1$.
The idea of such measurements was put forward many years
ago~\cite{Altarelli:67, Vainshtein:71}.
Several experiments in search of direct production of $C$-even resonances
in $e^+e^-$ collisions were performed, and very low upper limits on the
leptonic widths of $\eta^\prime$, $f_2(1270)$, $a_2(1320)$, and
$X(3872)$ mesons were set~\cite{snd1, kmd-3, snd2, besIII}.

The explanation of the smallness of the leptonic widths of $C$-even resonances is that
corresponding decays proceed via two virtual photons and therefore are suppressed by a factor
of $\alpha^4$, where $\alpha$ is the fine structure constant.

In this paper we consider $1^{++}$ meson $f_1(1285)$, its decay into
the $e^+e^-$ pair, and its direct production in $e^+e^-$ collisions.
The process $e^+e^-\to f_1 \to mesons$ is still not measured and
may be studied at the VEPP-2000 $e^+e^-$ collider
in experiments with the SND and CMD-3 detectors.

There is a quite extensive list of literature on the production of $1^{++}$
resonances in $e^+e^-$ annihilation. The direct production of $1^{++}$
states through the neutral current was evaluated many years ago in the
nonrelativistic quarkonium model~\cite{Kaplan:78}. The calculation of
the width $\Gamma(\chi_1 \to e^+e^-)$ was performed in the quarkonium and
vector meson dominance models (VMD)~\cite{Kuhn:79}. There are also some recent papers
devoted to $X(3872)$ and $\chi_{c1}$ decays into the $e^+e^-$ pair and their
production in $e^+e^-$ collisions (see~\cite{Denig:14, Czyz:16, Czyz:17, Kivel:16}
and references therein).
The production of $1^{++}$ resonances $R$ in two-photon collisions
($e^+e^-\to e^+e^-R$) was also extensively studied both
theoretically~\cite{Kopp:74, Renard:84, Cahn:87,
Schuler:98} and experimentally~\cite{Gidal:87, Aihara:88, Achard:02}.

The paper is organized as follows. In Sec.~\ref{sec:estimate} a simple
estimate of the width $\Gamma(f_1\to e^+e^-)$ is given.
In Sec.~\ref{sec:form_factors} we discuss the amplitude of
the $f_1 \to \gamma^*\gamma^*$ transition in a model-independent way.
In Sec.~\ref{sec:constants} amplitudes and coupling constants describing
the $f_1\to \rho^0\gamma$ decay are studied and constrained using
experimental data.
Section~\ref{sec:model} describes a choice of $f_1 \to \gamma^*\gamma^*$
form factors and the calculation of $\Gamma(f_1\to e^+e^-)$.
In Sec.~\ref{sec:cross_section} we estimate the $e^+e^-\to f_1 \to \eta \pi \pi$ cross section.
In Sec.~\ref{sec:asymmetry} charge asymmetry in the $e^+e^-\to \eta \pi^+ \pi^-$ process is studied.
And finally, in Sec.~\ref{sec:conclusion} we conclude.

\section{Simple estimate of $f_1\to e^+e^-$ decay width}
\label{sec:estimate}

\begin{figure}[h]
\includegraphics[scale=1.0]{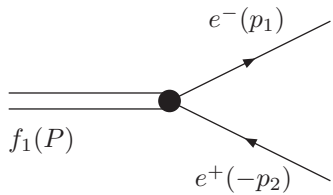}
\caption {The Feynman tree diagram for the $f_1 \to e^+e^-$ decay.}
\label{fig:1}
\end{figure}

It is convenient to start our discussion with the simple analysis of
the $f_1 \to e^+e^-$ decay (see the tree diagram in Fig.~\ref{fig:1}).

The electron and positron produced in this decay are ultrarelativistic in the $f_1$ rest frame.
So, in this frame $e^+$ and $e^-$ can be considered as massless and having
the definite helicities. An additional argument for neglecting the electron mass can be given.
In Ref.~\cite{Yang:12} the width of the $\chi_{c1}\to l^+l^-$ decay
was calculated with the finite lepton mass, and it was found that the mass effects are negligible.

\begin{figure}[h]
\includegraphics[scale=1.0]{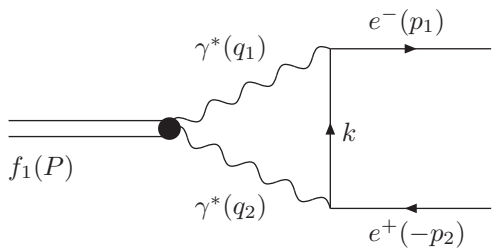}
\caption {One-loop diagram of the $f_1 \to e^+e^-$ decay with two intermediate
photons.}
\label{fig:2}
\end{figure}

To construct the decay amplitude, we notice that the decay
$e^+$ and $e^-$ may in principle be produced in {\it two} polarization states,
with the same ($j_z=0$) or opposite ($j_z=\pm 1$) helicities.
Here $j_z$ is the projection of the total $e^+e^-$ angular momentum $j$
onto the $z$ axis, which is directed along the $e^-$ momentum in the
$f_1$ rest frame. Because of conservation laws (in particular, conservation of $P$
and $C$ parities) only one polarization state with opposite helicities of
$e^+$ and $e^-$ is realized.
Therefore, there is only one $P$- and $C$-even invariant amplitude for
the $f_1 \to e^+e^-$ decay, which is written as
\begin{equation} \label{Mfee}
M(f_1 \to e^+e^-) = F_A \alpha^2 \widetilde{e}_\mu \bar{u}\gamma^\mu\gamma^5 v,
\end{equation}
where $\widetilde{e}_\mu$ is the $C$-even axial vector describing the $f_1$ meson,
$\bar{u}\gamma^\mu\gamma^5 v = j_A^\mu$ is the axial current, and $F_A$ is the
dimensionless coupling constant. Since $f_1$ meson is $C$-even, it decays
into $e^+e^-$ via two virtual photons as depicted in Fig.~\ref{fig:2}.
This explains the origin of the factor $\alpha^2$ in (\ref{Mfee}).

Using the amplitude (\ref{Mfee}) it is easy to calculate the decay width
\begin{equation}
\Gamma(f_1 \to e^+e^-) = \frac{\alpha^4|F_A|^2}{12\pi}m_f,
\end{equation}
where $m_f$ is the $f_1$ mass, $m_f = 1282.0$ MeV~\cite{pdg16}.

For a naive estimate, it is natural to assume that the coupling constant $F_A$
is of the order of unity, $|F_A|\sim 1$. In (\ref{Mfee}) we have
already written explicitly the small factor $\alpha^2$, and there are not any
additional small factors. So, we obtain that $\Gamma(f_1 \to e^+e^-)\sim 0.1$ eV.

In what follows we calculate this width in a certain model and find that this
simple estimate is correct by the order of magnitude.

\section{Model-independent description of $f_1 \to e^+e^-$ amplitude}
\label{sec:form_factors}

To calculate the width $\Gamma(f_1 \to e^+e^-)$ more accurately,
we should know the amplitude of the $f_1 \to \gamma^*\gamma^*$ transition
(see Fig.~\ref{fig:2}). This amplitude must be symmetric with respect to
the permutation of virtual photons and must vanish when both photons are on shell
($f_1 \to \gamma\gamma$ decay is forbidden by the Landau-Yang
theorem~\cite{Landau:48}).
Also the amplitude of the $f_1 \to \gamma^*\gamma^*$ transition must
contain {\it two} independent terms, corresponding to {\it two} different 
polarization states in the $f_1$ rest frame.
These states can be denoted as $TT$ (when both virtual photons are transversal)
and $TL$ (when the first photon is transversal and the second photon is longitudinal).
States $TL$ and $LT$ are the same here due to the photon identity.
The polarization state $LL$ (when both virtual photons are longitudinal) does
not exist, because the $f_1$ meson is axial one.

So, the $f_1 \to \gamma^*\gamma^*$ amplitude is parametrized in general
by two dimensionless form factors, $F_1(q_1^2, q_2^2)$ and $F_2(q_1^2, q_2^2)$,
which are functions of photon momenta squared. We choose this amplitude in
the following form based on amplitudes used, e.g., in Refs.~\cite{Kuhn:79, Kopp:74, Renard:84}:
\begin{multline} \label{Mfgg}
M(f_1 \to \gamma^*\gamma^*) = \frac{\alpha}{m^2_f}
F_1(q_1^2, q_2^2) i\epsilon_{\mu\nu\rho\sigma}q_1^\mu e_1^{*\nu} q_2^\rho e_2^{*\sigma}
\widetilde{e}^\tau(q_1-q_2)_\tau +\\
+ \frac{\alpha}{m^2_f} \left\{F_2(q_1^2, q_2^2) i\epsilon_{\mu\nu\rho\sigma}q_1^\mu e_1^{*\nu}
\widetilde{e}^\rho \left[q_2^\sigma e_2^{*\lambda} q_{2\lambda}-e_2^{*\sigma} q_2^2 \right] +
F_2(q_2^2, q_1^2) i\epsilon_{\mu\nu\rho\sigma}q_2^\mu e_2^{*\nu}
\widetilde{e}^\rho \left[q_1^\sigma e_1^{*\lambda} q_{1\lambda}-e_1^{*\sigma} q_1^2 \right]\right\},
\end{multline}
where $e_1$, $e_2$, and $\widetilde{e}$ are the polarization vectors of
the first photon, second photon, and $f_1$ meson, respectively.
In this expression the form factor $F_1$ corresponds to transversal
photons ($TT$), and the form factor $F_2$ describes a combination of
$TT$ and $LT$ polarization states.

Because of the Bose symmetry form factor $F_1(q_1^2, q_2^2)$ must be
antisymmetric, $F_1(q_1^2, q_2^2)=-F_1(q_2^2, q_1^2)$.
As it should be, the amplitude~(\ref{Mfgg}) vanishes when both photons are
on shell. Indeed, the first term vanishes because of $F_1(0, 0)=0$,
while all terms in the last line of (\ref{Mfgg})
vanish because $q^2=0$ and $e^\lambda q_\lambda=0$ for real photons.

We substitute this $f_1 \to \gamma^*\gamma^*$ amplitude into
the expression for the one-loop diagram (see Fig.~\ref{fig:2}) and perform
straightforward calculation in the Feynman gauge, using the identity
\begin{equation} \label{identity}
i\epsilon_{\mu\nu\rho\sigma} \gamma^\sigma =
\left(\gamma_\mu \gamma_\nu \gamma_\rho - g_{\mu\nu}\gamma_\rho + g_{\mu\rho}\gamma_\nu - g_{\nu\rho}\gamma_\mu \right)\gamma^5,
\end{equation}
and Dirac equations for massless electron and positron, $\bar{u}\hat{p}_1=0$ and $\hat{p}_2v=0$.
This leads to the following expression for the $f_1 \to e^+e^-$ amplitude:
\begin{multline} \label{Mfeeint}
M(f_1 \to e^+e^-) = -\frac{16\pi i\alpha^2}{m^2_f} \widetilde{e}^\mu P^\nu \bar{u}\gamma^\lambda \gamma^5v
\int \frac{d^4 k}{(2\pi)^4}\frac{k_\mu k_\nu k_\lambda}{k^2 q_1^2 q_2^2} F_1(q_1^2, q_2^2)-\\
- \frac{8\pi i\alpha^2}{m^2_f} \widetilde{e}^\mu \bar{u}\gamma^\nu \gamma^5 v
\int \frac{d^4 k}{(2\pi)^4}\frac{k_\mu k_\nu}{k^2 q_1^2 q_2^2}
\left\{F_2(q_1^2, q_2^2)q_2^2 + F_2(q_2^2, q_1^2)q_1^2 \right\}+\\
+ \frac{4\pi i\alpha^2}{m^2_f} \widetilde{e}_\mu \bar{u}\gamma^\mu \gamma^5 v
\int \frac{d^4 k}{(2\pi)^4}\frac{1}{k^2 q_1^2 q_2^2}
\left\{F_2(q_1^2, q_2^2)\left[k^2(p_1p_2+p_1k-p_2k)-2q_2^2(p_1k)+2q_2^2k^2\right]+\right.\\
\left. + F_2(q_2^2, q_1^2) \left[k^2(p_1p_2+p_1k-p_2k)+2q_1^2(p_2k)+2q_1^2k^2\right] \right\},
\end{multline}
where $q_1=p_1-k$ and $q_2=p_2+k$.

\section{Constants of $f_1\to \rho^0\gamma$ decay from experimental data} \label{sec:constants}

One cannot calculate the width $\Gamma(f_1 \to e^+e^-)$ in a model-independent
way, because the explicit form of functions $F_1$ and $F_2$ in (\ref{Mfeeint}) is
unknown. So, we have to choose some reasonable model.

\begin{figure}[h]
\includegraphics[scale=1.0]{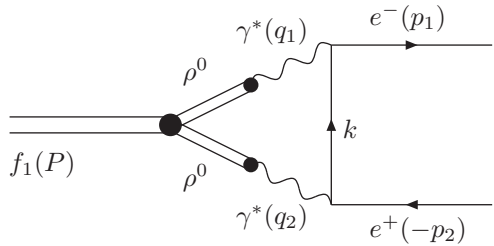}
\caption {The VMD mechanism of the $f_1 \to e^+e^-$ decay with two
intermediate $\rho^0$ mesons.}
\label{fig:3}
\end{figure}

We assume that the main contribution to the amplitude $M(f_1 \to e^+e^-)$ 
comes from the diagram depicted in Fig.~\ref{fig:3}, where both virtual photons are coupled
with the $f_1$ meson via intermediate $\rho^0$ mesons.
However, we do not take into account here direct $f_1 \gamma^*  \gamma^*$,
$f_1 \rho^0 \gamma^*$, and $f_1  \phi \gamma^*$ couplings.
One of the arguments is that dimensional analysis shows that form factors $F_1$ and $F_2$ 
should decrease rapidly with increasing momentum $k$ in order to avoid divergences in (3.3).
Even if form factors $F_1$ and $F_2$ behave as $1/k^2$ (it corresponds to 
$f_1 \rho^0 \gamma^*$ or $f_1  \phi \gamma^*$ couplings), then the amplitude (3.3) 
diverges logarithmically. 
This is the hint that both virtual photons couple with the $f_1$ meson via some massive vector mesons.
In such a case form factors $F_1$ and $F_2$ behave as $1/k^4$ and the amplitude (3.3) 
does not diverge.
Experimental data show that one of the main $f_1$ decay channels, $f_1 \to 4\pi$ 
[$\mathcal{B}(f_1\to 4\pi)\approx 33\%$], proceeds mainly via the intermediate $\rho\rho$
state~\cite{Barberis:00}. Other evidence of this mechanism is a large (5.5\%)
branching ratio of radiative $f_1 \to \rho^0\gamma$ decay~\cite{pdg16}.
So, the assumption that $f_1 \rho^0 \rho^0$ coupling
gives the main contribution to the amplitude $M(f_1 \to e^+e^-)$
looks quite reasonable. 

The amplitude $f_1 \to \rho^{0*} \rho^{0*}$ can be written by analogy
with (\ref{Mfgg}),
\begin{multline} \label{Mfrhorho}
M(f_1 \to \rho^{0*} \rho^{0*}) = \frac{1}{m^2_f}
h_1(q_1^2, q_2^2) i\epsilon_{\mu\nu\rho\sigma}q_1^\mu e_1^{*\nu} q_2^\rho e_2^{*\sigma}
\widetilde{e}^\tau(q_1-q_2)_\tau +\\
+ \frac{1}{m^2_f} \left\{h_2(q_1^2, q_2^2) i\epsilon_{\mu\nu\rho\sigma}q_1^\mu e_1^{*\nu}
\widetilde{e}^\rho \left[q_2^\sigma e_2^{*\lambda} q_{2\lambda}-e_2^{*\sigma} q_2^2 \right] +
h_2(q_2^2, q_1^2) i\epsilon_{\mu\nu\rho\sigma}q_2^\mu e_2^{*\nu}
\widetilde{e}^\rho \left[q_1^\sigma e_1^{*\lambda} q_{1\lambda}-e_1^{*\sigma} q_1^2 \right]\right\}.
\end{multline}

Some parameters of the model can be constrained from experimental data on
$f_1 \to \rho^0\gamma$ decay.
The corresponding amplitude can be obtained from (\ref{Mfgg}),
where all particles should be considered on shell. 
For $q_1^2=m_\rho^2$ 
(here $m_\rho=775.26$ MeV~\cite{pdg16} is the $\rho^0$ mass), 
$q_2^2=0$, and $e_1 q_1 = e_2 q_2 = 0$ we obtain
\begin{equation} \label{Mfrhogam}
M(f_1 \to \rho^0\gamma) = \frac{\alpha}{m^2_f}
g_1 i\epsilon_{\mu\nu\rho\sigma}p^\mu \epsilon^{*\nu} q^\rho e^{*\sigma}
\widetilde{e}^\tau(p-q)_\tau - \frac{\alpha m^2_\rho}{m^2_f}g_2 i\epsilon_{\mu\nu\rho\sigma}\widetilde{e}^\mu \epsilon^{*\nu} q^\rho e^{*\sigma},
\end{equation}
where $e, \epsilon$, and $\widetilde{e}$ are polarization vectors of photon, $\rho^0$, and $f_1$, respectively; 
$p$ and $q$ are momenta of $\rho^0$ and photon. 
This amplitude contains two complex coupling constants, $g_1$ and
$g_2$, because there are two different polarization states.
The first state is when $\rho^0$ meson polarization is longitudinal ($L$) in the $f_1$ rest
frame, and the second one is when $\rho^0$ meson polarization is transversal ($T$).
In the expression (\ref{Mfrhogam}) the coupling constant $g_1$ corresponds to
the $T$ polarization state of $\rho^0$, and $g_2$ corresponds to 
a combination of $L$ and $T$ polarization states.
In the $f_1$ rest frame the ratio of the longitudinal and transversal parts 
of this amplitude is the following:
\begin{equation} \label{MLMT}
\frac{M_L(f_1 \to \rho^0\gamma)}{M_T(f_1 \to \rho^0\gamma)}=\frac{\sqrt{\xi}g_2}{(1-\xi)g_1+\xi g_2},
\end{equation} 
where $\xi=m^2_\rho/m^2_f \approx 0.37$.

Now it is straightforward to calculate the width of $f_1 \to \rho^0\gamma$ decay,
\begin{equation} \label{frhogamma}
\Gamma(f_1 \to \rho^0\gamma)= \frac{\alpha^2}{96\pi}m_f (1-\xi)^3
\left[(1-\xi)^2|g_1|^2+\xi(1+\xi)|g_2|^2+2\xi(1-\xi)|g_1||g_2|\cos{\delta}\right].
\end{equation}
Since the parameters $g_1$ and $g_2$
do not correspond to different polarization states
[see the comment after Eq. (\ref{Mfgg})],
the interference term does not vanish after summation over polarizations, and
expression (\ref{frhogamma}) contains $\delta=\phi_1-\phi_2$, which is the relative
phase of the complex constants $g_1$ and $g_2$.

The expression (\ref{frhogamma}) represents one relation between three
unknown parameters $g_1$, $g_2$, and $\delta$.
One more relation can be derived from the polarization experiments.
The ratio of the contributions of two $\rho^0$ helicity states,
$r=\rho_{LL}/\rho_{TT}=3.9\pm 0.9 \pm 1.0$, was determined in the VES
experiment~\cite{Amelin:95}
from the analysis of angular distributions in the reaction $f_1
\to \rho^0\gamma \to \pi^+\pi^-\gamma$,
\begin{equation} \label{frgppg}
|M(f_1 \to \rho^0\gamma \to \pi^+\pi^-\gamma)|^2 \sim
\rho_{LL}\cos^2{\theta}+\rho_{TT}\sin^2{\theta},
\end{equation}
where $\rho_{LL}$ and $\rho_{TT}$ are density matrix elements corresponding
to longitudinal and transverse $\rho^0$ mesons, respectively;
$\theta$ is the angle between $\pi^+$ and $\gamma$ momenta in
the $\rho^0$ rest frame. Integrating over $d(\cos{\theta})$, one easily finds that
\begin{equation} \label{Gfrgppg}
\Gamma(f_1 \to \rho^0\gamma \to \pi^+\pi^-\gamma) \sim \rho_{LL}+2\rho_{TT}.
\end{equation}

Calculation of $|M(f_1 \to \rho^0\gamma \to \pi^+\pi^-\gamma)|^2$ with the
amplitude (\ref{Mfrhogam}) leads to the following
ratio of the coefficients at $\cos^2{\theta}$ and $\sin^2{\theta}$:
\begin{equation} \label{condition2}
r=\frac{2\xi|g_2|^2}{(1-\xi)^2|g_1|^2+\xi^2|g_2|^2+2\xi(1-\xi)|g_1||g_2|\cos{\delta}},
\end{equation}
which equals to $\rho_{LL}/\rho_{TT}$ from Ref.~\cite{Amelin:95}.

The decay width (\ref{frhogamma}) can be
presented as a sum of contributions of different polarization states,
\begin{equation}
\Gamma(f_1 \to \rho^0\gamma)= \frac{\alpha^2}{96\pi}m_f (1-\xi)^3
\left[g_{TT} + g_{LL} \right],
\end{equation}
where
\begin{equation} \label{gTT}
g_{TT}=(1-\xi)^2|g_1|^2+\xi^2|g_2|^2+2\xi(1-\xi)|g_1||g_2|\cos{\delta},
\end{equation}
\begin{equation} \label{gLL}
g_{LL}=\xi|g_2|^2.
\end{equation}
Note that $g_{LL}/g_{TT}=r/2=\rho_{LL}/2\rho_{TT}$ in agreement with (\ref{Gfrgppg}).

Recently CLAS Collaboration at Jefferson Laboratory (JLab) published the results of 
the first measurements of the $f_1$ meson
in the photoproduction reaction $\gamma p \to f_1 p$ off a proton target~\cite{Dickson:16}.
The first estimate of the cross section of this reaction in the JLab kinematics and 
suggestion that its measurement with the CLAS detector is possible were reported in Ref.~\cite{Kochelev:09}.

According to the data of CLAS Collaboration, the branching ratio $\mathcal{B}(f_1 \to \rho^0\gamma)$
equals $(2.5\pm0.9)\%$, which is substantially smaller than the PDG value, $(5.5\pm1.3)\%$.
A few theoretical models consistent with this result are proposed already~\cite{Wang:17, Wang:2017, Osipov:17}.
The total $f_1$ width measured by CLAS Collaboration $\Gamma_f^{CLAS}=(18.4\pm1.4)$ MeV 
is also considerably smaller than the PDG value, $\Gamma_f=(24.1\pm1.0)$ MeV.
Therefore, below we present the results of our calculations for PDG averages
and CLAS Collaboration values as well.

Using the experimental result $r=3.9\pm 0.9 \pm 1.0$~\cite{Amelin:95} we obtain
\begin{equation} \label{g_TT}
\alpha^2 g_{TT}=\frac{96\pi \mathcal{B}(f_1\to \rho^0\gamma)\Gamma_f}{m_f (1-m^2_\rho/m^2_f)^3}\frac{2}{r+2}
= 0.41 \pm 0.14, \hspace{10mm} \alpha^2 g_{TT}^{CLAS}=0.14 \pm 0.06,
\end{equation}
\begin{equation} \label{g_LL}
\alpha^2 g_{LL}=\frac{96\pi \mathcal{B}(f_1\to \rho^0\gamma)\Gamma_f}{m_f (1-m^2_\rho/m^2_f)^3}\frac{r}{r+2}
= 0.81 \pm 0.22, \hspace{10mm} \alpha^2 g_{LL}^{CLAS}=0.28 \pm 0.18,
\end{equation}
and find the magnitude of coupling constant $g_2$,
\begin{equation} \label{g_2}
\alpha|g_2| = 1.5 \pm 0.2, \hspace{10mm}  \alpha|g_2|^{CLAS}=0.87 \pm 0.18.
\end{equation}
It is seen that values of this coupling constant obtained for PDG and CLAS data are essentially different.

It is impossible to extract the magnitude of
the constant $g_1$ and/or the phase $\delta$ from the experimental data. 
From Eq.~(\ref{gTT}) we obtain the following relation 
between the absolute value of $g_1$ and two other parameters:
\begin{equation} \label{g_1}
|g_1|=\frac{1}{1-\xi}\left(-\xi|g_2|\cos{\delta}+\sqrt{(\xi|g_2|\cos{\delta})^2+g_{TT}-\xi g_{LL}}\right).
\end{equation}
Taking into account that $-1\leq\cos{\delta}\leq 1$, we obtain
\begin{equation} \label{g1}
0.16 \lesssim \alpha|g_1|\lesssim 1.9, \hspace{10mm} 0.09 \lesssim \alpha|g_1|^{CLAS}\lesssim 1.1,
\end{equation}
for the central values of $g_{TT}$, $g_{LL}$, and $|g_2|$; see (\ref{g_TT}),
(\ref{g_LL}), and (\ref{g_2}).
The upper and lower limits on $|g_1|$ correspond to $\delta=\pi$ and $\delta=0$, respectively.

It is seen that there is a large uncertainty in the value of $|g_1|$.
Indeed, $|g_1|$ could be of the same order of magnitude as $|g_2|$, if $\delta$ is close to $\pi$, 
and of the order of magnitude smaller, if $\delta$ is close to 0.
Moreover, quite large experimental uncertainties in the polarization
experiment~\cite{Amelin:95} allow one to speculate that $|g_1|$ could be very small or even negligible.
Indeed, in the case $g_1=0$ one obtains from (\ref{condition2}) that $r=2/\xi \approx 5.5$,
which is not very far from the central value $r=3.9$.

There are some papers concerning $f_1 \to \rho^0\gamma$ decay,
where the amplitude $M(f_1 \to \rho^0\gamma)$ is parametrized only 
by one constant~\cite{Kochelev:09, Wang:2017, Osipov:17}.
Therein the ratio of $M_L(f_1 \to \rho^0\gamma)$ to $M_T(f_1 \to \rho^0\gamma)$ 
is equal to $m_f/m_{\rho}=1/\sqrt{\xi}$,
which coincides with the expression (\ref{MLMT}) at $g_1=0$.
So, these models correspond to our model at $g_1=0$.

There are also papers where the amplitude $M(f_1 \to \rho^0\gamma)$ is parametrized by two constants; see, e.g., Ref.~\cite{Lutz:08}. 
Therein two relations between parameters of $f_1 \to \rho^0\gamma$ decay are obtained
using $\Gamma(f_1 \to \rho^0\gamma)$ and $r=\rho_{LL}/\rho_{TT}$.
In this paper the $f_1$ meson is considered as the molecular state and
$f_1 \to \rho^0\gamma$ decay is studied in the chiral effective field theory.

\section{Calculation of $f_1\to e^+e^-$ decay width} \label{sec:model}

In order to construct the amplitude $M(f_1 \to e^+e^-)$ (see Fig.~\ref{fig:3})
we consider also the transition $\rho^{0*} \to \gamma^*$.
The Lagrangian of such a transition in gauge-invariant form reads
\begin{equation}
\mathcal{L}=\frac{1}{2}g_{\rho\gamma}V^{\mu\nu}F_{\mu\nu},
\end{equation}
where $g_{\rho\gamma}$ is the coupling constant, and $V_{\mu\nu}$ and $F_{\mu\nu}$
are the tensors of the $\rho^0$ meson field and electromagnetic field,
respectively.
Therefore, we obtain the corresponding amplitude,
\begin{equation}
M(\rho^0 \to \gamma)=g_{\rho\gamma}(q^2 g_{\mu\nu}-q_\mu q_\nu) \epsilon^\mu e^{*\nu},
\end{equation}
where  $q$ is the momentum, and $\epsilon$ and $e$ are polarization vectors of
the $\rho^0$ meson and photon, respectively.
The coupling constant $g_{\rho\gamma}$
can be found from $\rho^0 \to e^+e^-$ decay width,
\begin{equation}
\Gamma(\rho^0 \to e^+e^-)=\frac{1}{3}\alpha g_{\rho\gamma}^2 m_\rho.
\end{equation}
Using the experimental values $\Gamma(\rho^0 \to e^+e^-)\approx 6.98$ keV and
$m_\rho=775.26$ MeV \cite{pdg16} we derive $g_{\rho\gamma}\approx 0.06$.

Now we can compare the amplitudes described by diagrams in Figs.~\ref{fig:2}
and \ref{fig:3} and obtain the relation between form factors $F_1$ and $h_1$, $F_2$ and $h_2$,
\begin{equation} \label{Fh}
\alpha F_{1,2} (q_1^2, q_2^2) = \frac{g^2_{\rho\gamma} q_1^2 q_2^2}
{(q_1^2-m_\rho^2+i m_\rho \Gamma_\rho)(q_2^2-m_\rho^2+i m_\rho \Gamma_\rho)} h_{1,2} (q_1^2, q_2^2),
\end{equation}
where $\Gamma_\rho = 147.8$ MeV is the $\rho^0$-meson width.
In what follows we consider $\Gamma_\rho$ as a constant parameter, because 
to account for its dependence on the momentum squared,
$\Gamma_\rho (q^2)$, seems to be beyond the accuracy of our calculation.

Comparison (\ref{Mfrhorho}) with (\ref{Mfrhogam}) leads to relations
\begin{equation} \label{h1}
\lim_{q_2^2 \to 0} q_2^2 h_1 (m_\rho^2, q_2^2) = -\frac{\alpha g_1}{g_{\rho\gamma}} \left(m_\rho^2-i m_\rho\Gamma_\rho \right),
\end{equation}
\begin{equation} \label{h2}
\lim_{q_2^2 \to 0} q_2^2 h_2 (q_2^2, m_\rho^2) = -\frac{\alpha g_2}{g_{\rho\gamma}} \left(m_\rho^2-i m_\rho\Gamma_\rho \right).
\end{equation}

\begin{figure}[h]
\center
\begin{tabular}{c c c}
\includegraphics[scale=1.0]{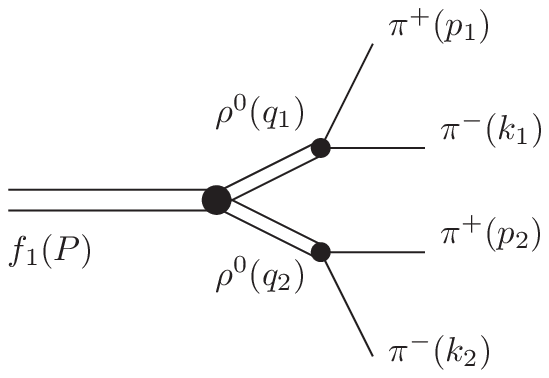} & &
\includegraphics[scale=1.0]{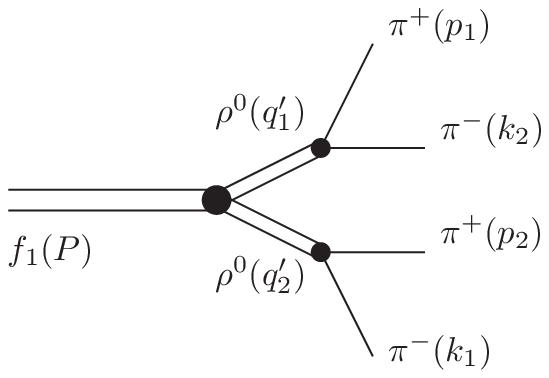}
\end{tabular}
\caption {Feynman diagrams of $f_1 \to \pi^+ \pi^- \pi^+ \pi^-$ decay.}
\label{fig:f2pi2pi}
\end{figure}

Now let us consider $f_1 \to \pi^+ \pi^- \pi^+ \pi^-$ decay.
Experimental data indicate that the main contribution to it 
is given by the intermediate state with two virtual $\rho$
mesons~\cite{Barberis:00} (see Fig.~\ref{fig:f2pi2pi}).
The vertex $f_1\rho\rho$ contains the form factors $h_1$ and $h_2$ of our model.
Certainly, these form factors should meet the requirements 
that the result of the calculation of the $f_1 \to \pi^+ \pi^- \pi^+ \pi^-$ decay width
should be in a good agreement with the experimental value. 

We parametrize the $\rho^0 \to \pi^+ \pi^-$ amplitude as
\begin{equation}
M(\rho^0 \to \pi^+ \pi^-) = i f_{\rho\pi\pi} e_\mu p^\mu,
\end{equation}
where $e_\mu$ is the polarization vector of the $\rho^0$ meson, 
and $p^\mu$ is the momentum of the $\pi^+$ meson.
We neglect the $q^2$ dependence of $f_{\rho\pi\pi}$ and obtain the following value:
\begin{equation} \label{frhopipi}
f_{\rho\pi\pi} = \left(\frac{192\pi \Gamma(\rho^0 \to \pi^+ \pi^-)}{m_\rho (1-4m_\pi^2/m_\rho^2)^{3/2}}\right)^{1/2}\approx 11.9,
\end{equation}
where $\Gamma(\rho^0 \to \pi^+ \pi^-) \approx \Gamma_\rho = 147.8$ MeV, and $m_\pi = 139.57$ MeV is the mass of $\pi^\pm$ mesons.

Taking into account that form factor $F_1(q_1^2, q_2^2)$ is antisymmetric,
and using relations (\ref{Fh}), (\ref{h1}), and (\ref{h2}) as a hint,
we write the form factors $F_1$ and $F_2$ as
\begin{equation} \label{F1}
F_1(q_1^2, q_2^2)=\frac{g_{\rho\gamma} g_1 (m_\rho^2 - i m_\rho \Gamma_\rho) (q_2^2-q_1^2)}{(q_1^2-m_\rho^2+i m_\rho \Gamma_\rho)(q_2^2-m_\rho^2+i m_\rho \Gamma_\rho)},
\end{equation}
\begin{equation} \label{F2}
F_2(q_1^2, q_2^2)=\frac{g_{\rho\gamma} g_2 (m_\rho^2 - i m_\rho \Gamma_\rho) (-m_\rho^2)}{(q_1^2-m_\rho^2+i m_\rho \Gamma_\rho)(q_2^2-m_\rho^2+i m_\rho \Gamma_\rho)}.
\end{equation}

\begin{figure}[h]
\center
\begin{tabular}{c c c}
\includegraphics[width=0.49\textwidth]{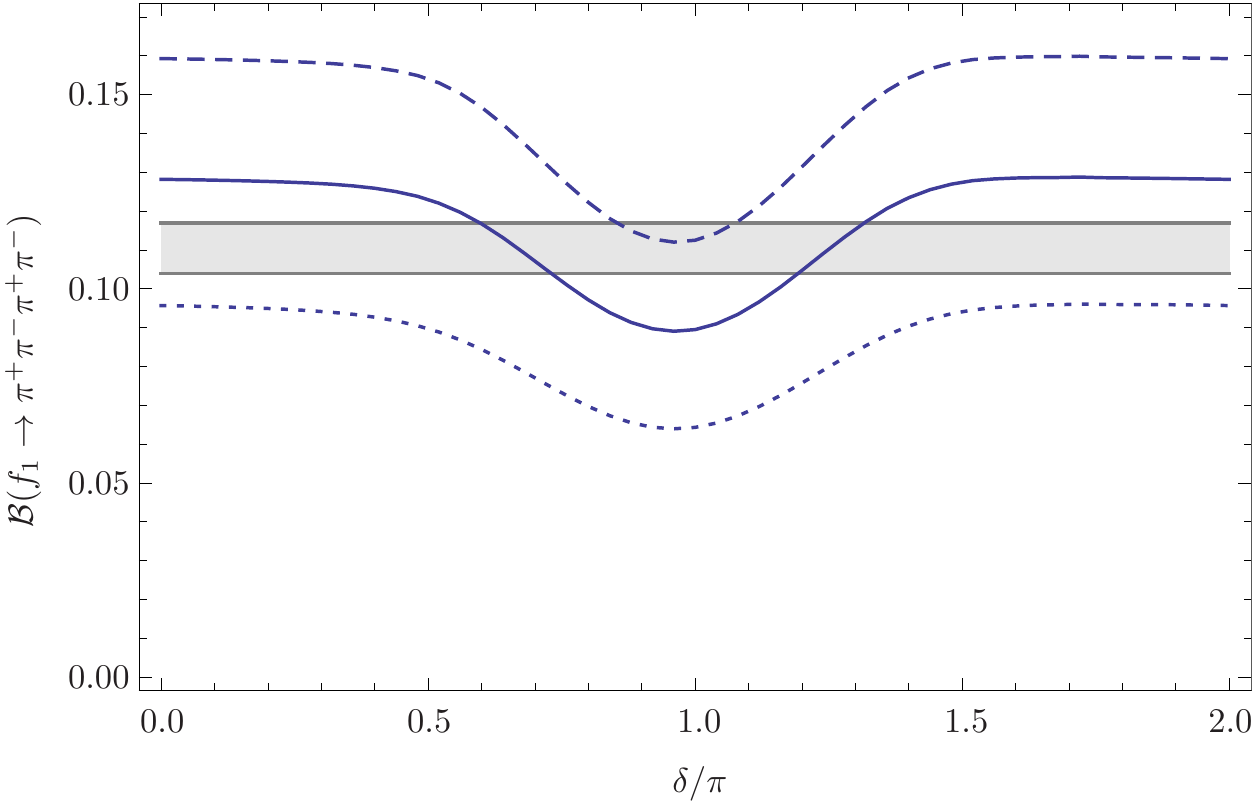} & &
\includegraphics[width=0.49\textwidth]{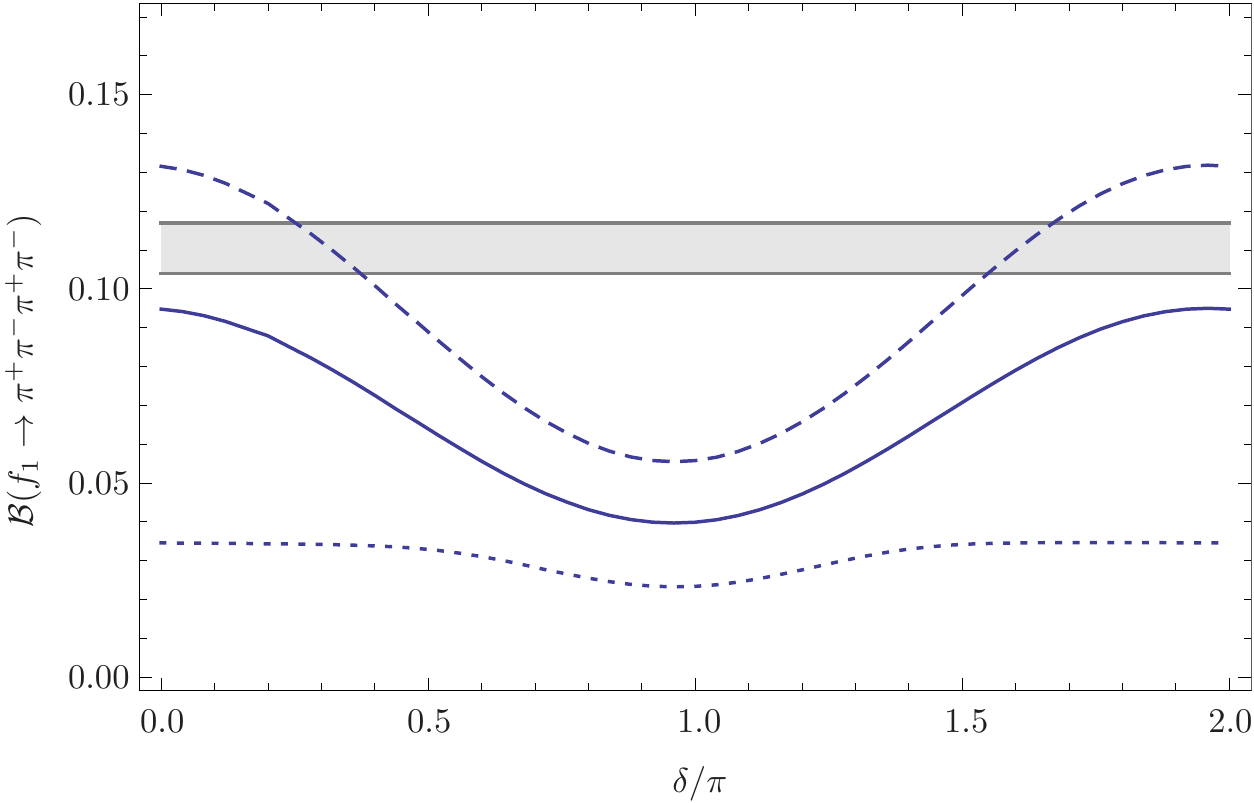}\\
PDG data & & CLAS Collaboration data
\end{tabular}
\caption {The branching ratio of the $f_1 \to \pi^+ \pi^- \pi^+ \pi^-$ decay
for the certain choice of the form factors $F_1 (q_1^2, q_2^2)$ and
$F_2 (q_1^2, q_2^2)$; see Eqs.~(\ref{F1}) and (\ref{F2}).
The solid line corresponds to the $f_1 \to \pi^+ \pi^- \pi^+ \pi^-$ branching ratio 
calculated using the central values: $r=3.9$ and
$\mathcal{B}(f_1\to \rho^0\gamma) = 5.5\%$
or $\mathcal{B}^{CLAS}(f_1\to \rho^0\gamma) = 2.5\%$.
Dashed and dotted lines indicate $1\sigma$ deviations for
$\mathcal{B}(f_1 \to \pi^+ \pi^- \pi^+ \pi^-)$.
The shaded horizontal band denotes the value allowed experimentally,
$\mathcal{B}(f_1 \to \pi^+ \pi^- \pi^+ \pi^-) = \left(11.0^{+0.7}_{-0.6}\right)\%$.}
\label{fig:Bf2pi2pi}
\end{figure}

The result of numerical calculation with corresponding form factors $h_1$ and $h_2$ is shown in
Fig.~\ref{fig:Bf2pi2pi}. The solid line depicts the branching ratio of
$f_1 \to \pi^+ \pi^- \pi^+ \pi^-$ decay, $\mathcal{B}_c$, calculating for the
central values: $r=\rho_{LL}/\rho_{TT}=3.9$ and $\mathcal{B}(f_1\to \rho^0\gamma) = 5.5\%$ 
or $\mathcal{B}^{CLAS}(f_1\to \rho^0\gamma) = 2.5\%$. 
Quite large experimental uncertainties $\Delta r\approx 1.3$ and $\Delta\mathcal{B}(f_1\to \rho^0\gamma)=1.3\%$
or $\Delta\mathcal{B}^{CLAS}(f_1\to \rho^0\gamma) = 0.9\%$
may lead to the substantial deviation of $\mathcal{B}(f_1 \to \pi^+ \pi^- \pi^+ \pi^-)$ 
from its central values $\mathcal{B}_c$. 
The results, corresponding to one standard deviation of
$\mathcal{B}$ from its central value, are shown in Fig.~\ref{fig:Bf2pi2pi} 
by dashed and dotted lines. The shaded horizontal band in Fig.~\ref{fig:Bf2pi2pi}
indicates values allowed experimentally, $\mathcal{B}(f_1 \to \pi^+ \pi^- \pi^+
\pi^-) = \left(11.0^{+0.7}_{-0.6}\right)\%$.

It is seen from Fig.~\ref{fig:Bf2pi2pi} that we still cannot derive the exact value
of the phase $\delta$ in our model because of large uncertainties of the model
parameters. Therefore, in what follows we treat $\delta$ as a free parameter.

To calculate the $f_1 \to e^+e^-$ branching ratio, we substitute the
expressions for $F_1 (q_1^2, q_2^2)$ and $F_2 (q_1^2, q_2^2)$
into (\ref{Mfeeint}) and perform the numerical calculations; then, comparing
the answer with (\ref{Mfee}), we obtain the following result for the
constant $F_A$:
\begin{equation}
F_A \simeq -\alpha g_1 \left(0.22+0.25i\right) - \alpha g_2 \left(0.75+0.57i\right).
\end{equation}
It is convenient to express complex numbers $g_1$ and $g_2$ in polar form as
$g_1=|g_1|\cdot e^{i\phi_1}$ and $g_2=|g_2|\cdot e^{i\phi_2}$, respectively.
Then using $\delta=\phi_1-\phi_2$ one can write the absolute square of the
constant $F_A$ as
\begin{equation} \label{F_A2}
|F_A|^2 \simeq \Big|e^{i\delta} \cdot \alpha |g_1| \cdot \left(0.22+0.25i\right) + \alpha |g_2| \cdot \left(0.75+0.57i\right)\Big|^2.
\end{equation}
Since $\alpha |g_2|\sim 1$ and $\alpha |g_1|\lesssim 1$ [see (\ref{g_2}) and (\ref{g1})], 
then $|F_A|\sim 1$, as expected.
In particular, for central values of $|g_1|$ and $|g_2|$ we get
$|F_A|\simeq 1.12$ for $\delta=0.7\pi$, $|F_A|\simeq 1.28$ for $\delta=1.3\pi$, 
and $|F_A|^{CLAS}\simeq 0.85$ for $\delta=0$.

\begin{figure}[h]
\center
\begin{tabular}{c c c}
\includegraphics[width=0.49\textwidth]{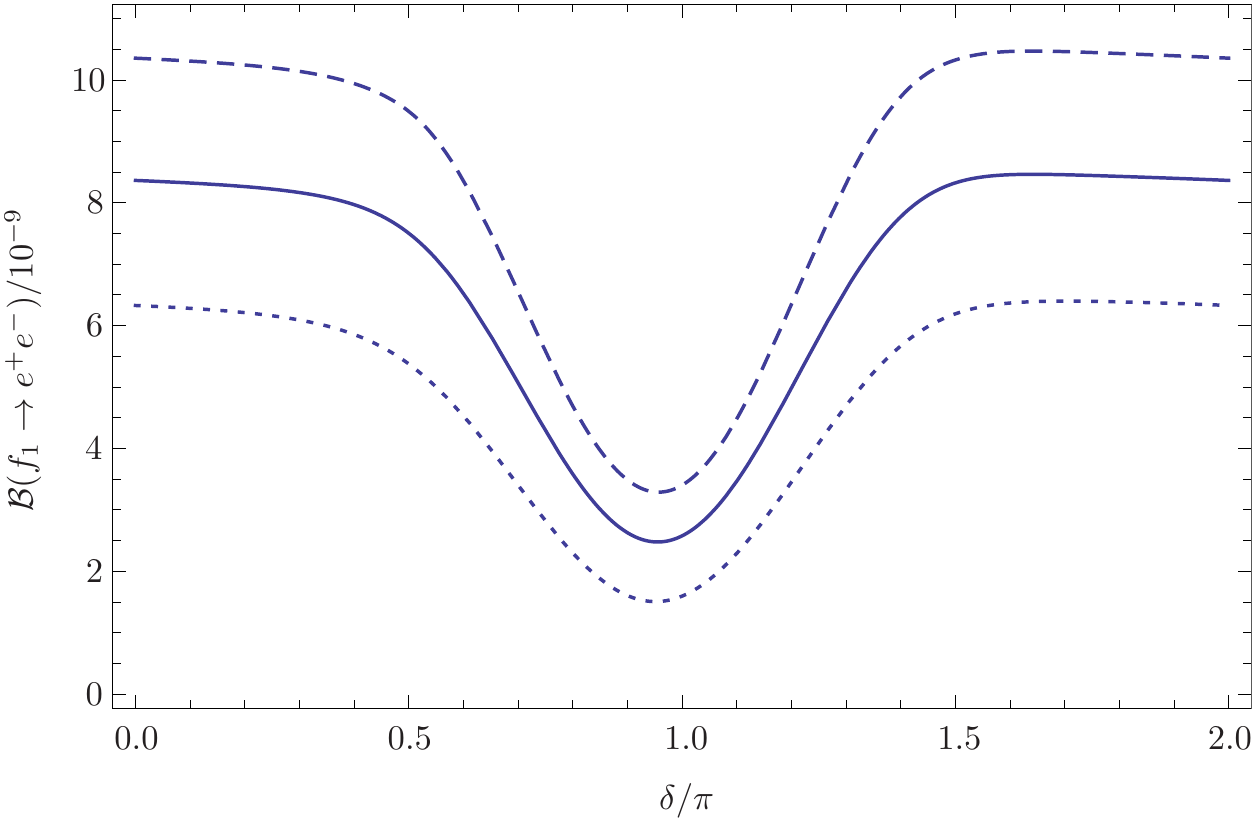} & &
\includegraphics[width=0.49\textwidth]{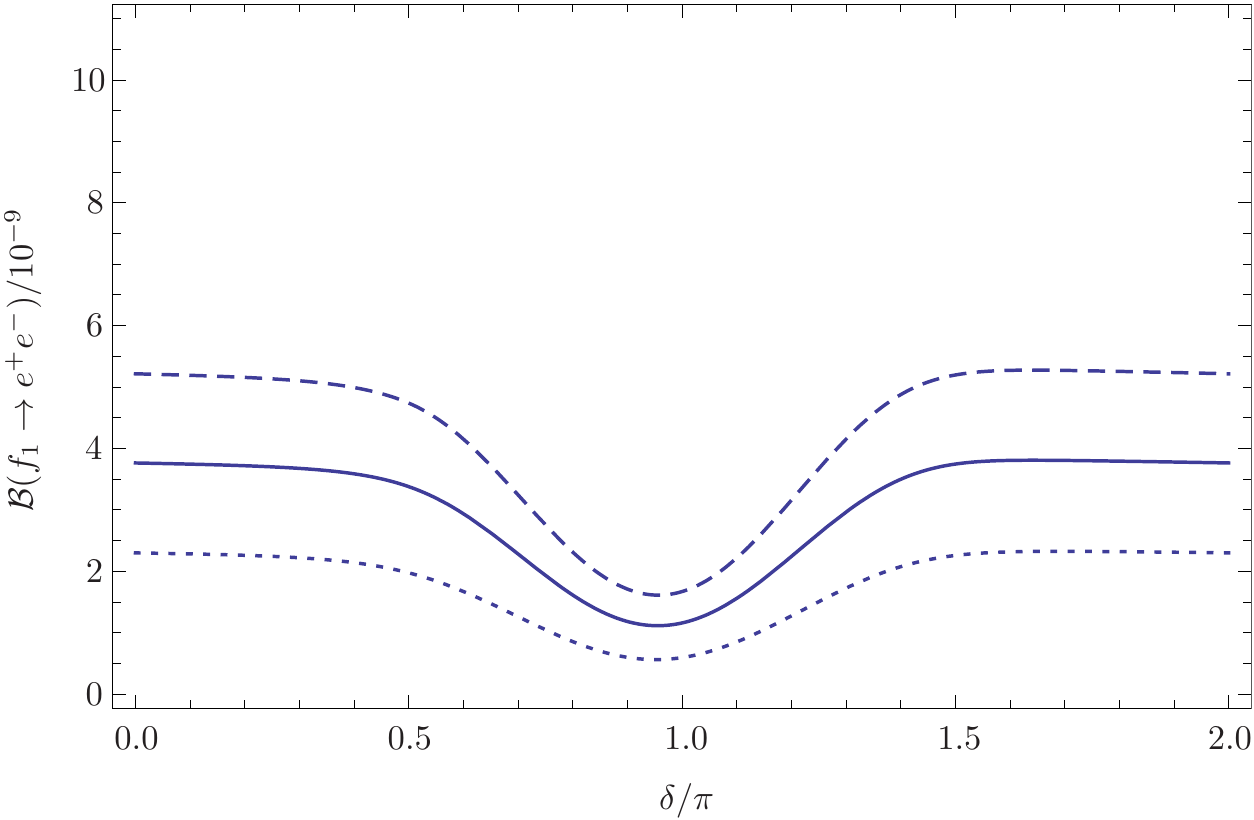}\\
PDG data & & CLAS Collaboration data
\end{tabular}
\caption {The branching ratio $\mathcal{B}(f_1 \to e^+e^-)$ as a function of
the relative phase $\delta$ in our model.
The solid line corresponds to $\mathcal{B}(f_1 \to e^+e^-)$
calculated for the central values: $r=3.9$ and
$\mathcal{B}(f_1\to \rho^0\gamma) = 5.5\%$ or
$\mathcal{B}^{CLAS}(f_1\to \rho^0\gamma) = 2.5\%$.
The dashed and dotted lines indicate $1\sigma$ deviations from
the $\mathcal{B}(f_1 \to e^+e^-)$ central value.}
\label{fig:Bfee}
\end{figure}

Now it is straightforward to calculate the branching ratio
$\mathcal{B}(f_1 \to e^+e^-)$ as a function of $\delta$.
Corresponding plots are shown in Fig.~\ref{fig:Bfee}, where the solid line
denotes $\mathcal{B}(f_1 \to e^+e^-)$
calculating for the central values: $r=3.9$ and
$\mathcal{B}(f_1\to \rho^0\gamma) = 5.5\%$
or $\mathcal{B}^{CLAS}(f_1\to \rho^0\gamma) = 2.5\%$.
Dashed and dotted lines indicate $1\sigma$ deviations.

We see that these functions are almost constant for $-\pi/2 < \delta < \pi/2$
and have a minimum near $\delta = \pi$. Such behavior can easily be
understood from (\ref{F_A2}) and (\ref{g_1}). Indeed, when $\cos{\delta}>0$,
the value of $|g_1|$ is quite small due to cancellation in (\ref{g_1}), so the main
contribution to $|F_A|^2$ is given by $|g_2|$ and the corresponding factor, which
are both independent of $\delta$. However, when $\delta$ is close to $\pi$ then $|g_1|$ is
comparable with $|g_2|$, so a quite strong cancellation occurs in (\ref{F_A2}), 
and therefore $\mathcal{B}(f_1 \to e^+e^-)$ is minimal.

It is seen from Figs.~\ref{fig:Bf2pi2pi} and \ref{fig:Bfee} that in our model  
the branching ratio $\mathcal{B}(f_1 \to e^+e^-)$ should be taken in the range
from $3 \cdot 10^{-9}$ for $\delta \simeq \pi$ to $8 \cdot 10^{-9}$ for
$\delta = 0$, and from $4 \cdot 10^{-9}$ for $\delta^{CLAS} = 0$ to $5 \cdot 10^{-9}$ for
$\delta^{CLAS} \simeq \pm 0.3\pi$,
\begin{equation} \label{Branch}
\mathcal{B}(f_1 \to e^+e^-) \simeq (3 \textendash 8) \cdot 10^{-9},
\hspace{10mm} \mathcal{B}^{CLAS}(f_1 \to e^+e^-) \simeq (4 \textendash 5) \cdot 10^{-9},
\end{equation}
and the corresponding decay width is
\begin{equation}
\Gamma (f_1 \to e^+e^-) \simeq 0.07 \textendash 0.19\ {\rm eV},
\hspace{10mm} \Gamma^{CLAS}(f_1 \to e^+e^-) \simeq 0.07 \textendash 0.10\ {\rm eV}.
\end{equation}
The values of the branching ratio and the decay width obtained for CLAS data lie in a more narrow interval
than the corresponding values obtained for PDG data.
However, both ranges of $\Gamma(f_1 \to e^+e^-)$ values are in good agreement with the
naive estimate $\Gamma\sim 0.1$ eV
(see the end of Sec.~\ref{sec:estimate}).

\section{Estimate of $e^+e^-\to f_1 \to \eta \pi \pi$ cross section}
\label{sec:cross_section}

\begin{figure}[h]
\center
\begin{tabular}{c c c}
\includegraphics[scale=1.0]{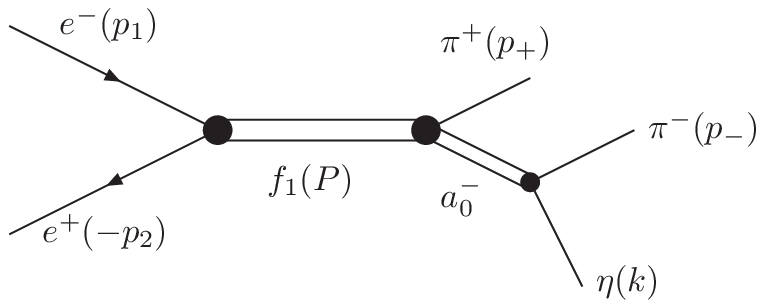} & &
\includegraphics[scale=1.0]{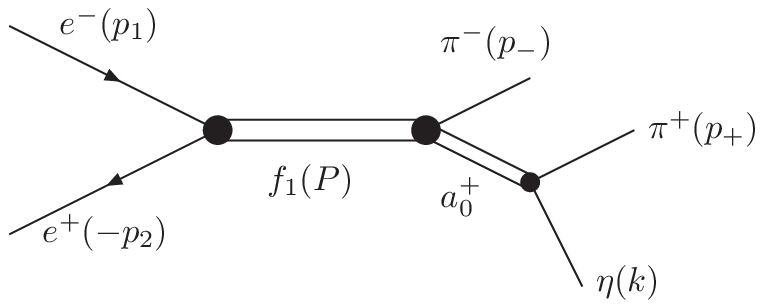}
\end{tabular}
\caption {The diagrams for $e^+e^-$ annihilation into the $\eta \pi^+ \pi^-$
final state via the intermediate $f_1$ and $a_0$ mesons.} \label{fig:eefa}
\end{figure}

\begin{figure}[h]
\includegraphics[scale=1.0]{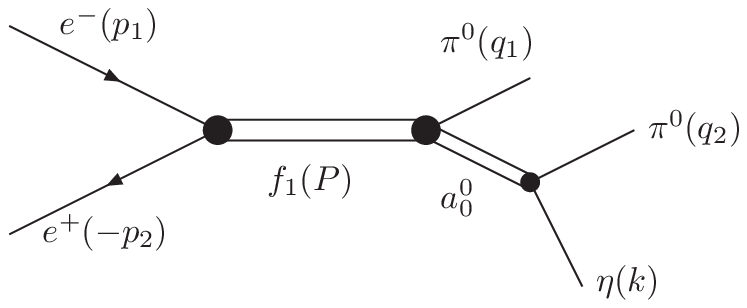}
\caption {The diagram for $e^+e^-$ annihilation into the $\eta \pi^0 \pi^0$
final state via the intermediate $f_1$ and $a_0$ mesons.} \label{fig:eefa3}
\end{figure}

Let us estimate the cross section of the process $e^+e^-\to f_1 \to \eta \pi \pi$,
which can be used for the study of\linebreak direct $f_1$ production in $e^+e^-$ collisions. 
Here, the $f_1 \to \eta \pi \pi$ decay proceeds mainly (approximately with 70\%
probability~\cite{pdg16}) through the intermediate $a_0(980)$ meson; see
Figs.~\ref{fig:eefa} and \ref{fig:eefa3}.

The branching ratio of the $f_1 \to a_0\pi$ decay is $(36\pm7)\%$~\cite{pdg16}.
Using the isospin symmetry we obtain
$\mathcal{B}(f_1\to a_0^+ \pi^-) = \mathcal{B}(f_1\to a_0^- \pi^+) = \mathcal{B}(f_1\to a_0^0 \pi^0)
= 1/3\,\mathcal{B}(f_1\to a_0 \pi) \approx 12\%$.

Since the $a_0$ meson is scalar and the $\pi$ meson is
pseudoscalar, the amplitude of the $f_1 \to a_0\pi$ decay can be written as
\begin{equation}
M(f_1 \to a_0\pi)=i g \chi_a^* \widetilde{\phi}_\pi^* \widetilde{e}_\mu p_\pi^\mu,
\end{equation}
where $g$ is the dimensionless coupling constant; 
$\chi_a$, $\widetilde{\phi}_\pi$, and $\widetilde{e}_\mu$
are wave functions of $a_0$, $\pi$ and $f_1$ mesons, respectively; 
and $p_\pi$ is the momentum of the $\pi$ meson.

The cross section of the $e^+e^-\to f_1\to a_0\pi$ process can easily be calculated,
\begin{equation}
\sigma(e^+e^-\to f_1\to a_0\pi)=\frac{12\pi}{m_f^2} \mathcal{B}(f_1 \to a_0\pi) \mathcal{B}(f_1 \to e^+e^-),
\end{equation}
where the center-of-mass energy equals the mass of the $f_1$ meson,
$\sqrt{s}=m_f$.
Using the experimental value for the branching ratio
$\mathcal{B}(f_1\to a_0\pi) = 0.36\pm0.07$ and
the result of our calculations (\ref{Branch}) for
$\mathcal{B}(f_1 \to e^+e^-)$,
we obtain
\begin{equation}
\sigma(e^+e^-\to f_1\to a_0\pi)\simeq 7.8 \textendash 30\ {\rm pb},
\hspace{10mm} \sigma^{CLAS}(e^+e^-\to f_1\to a_0\pi)\simeq 10 \textendash 20\ {\rm pb}.
\end{equation}

Assuming that the $a_0$ meson decays only into the $\eta \pi$ final state and 
using the relation
$\mathcal{B}(f_1 \to a_0^\pm \pi^\mp \to \eta \pi^+ \pi^-) = 2\,\mathcal{B}(f_1 \to a_0^0 \pi^0 \to \eta \pi^0 \pi^0) = 2/3\,\mathcal{B}(f_1 \to a_0\pi \to \eta \pi \pi)$,
we obtain the following estimates:
\begin{equation} \label{pi+pi-}
\sigma(e^+e^-\to f_1\to a_0^\pm \pi^\mp \to \eta \pi^+ \pi^-)\simeq 5.2 \textendash 20\ {\rm pb},
\hspace{3mm} \sigma^{CLAS}(e^+e^-\to f_1\to a_0^\pm \pi^\mp \to \eta \pi^+ \pi^-)\simeq 7 \textendash 13.3\ {\rm pb},
\end{equation}
\begin{equation} \label{pi0pi0}
\sigma(e^+e^-\to f_1\to a_0^0 \pi^0 \to \eta \pi^0 \pi^0)\simeq 2.6 \textendash 10\ {\rm pb},
\hspace{3mm} \sigma^{CLAS}(e^+e^-\to f_1\to a_0^0 \pi^0 \to \eta \pi^0 \pi^0)\simeq 3.5 \textendash 6.7\ {\rm pb}.
\end{equation}
It is seen that the values of cross sections obtained for PDG and CLAS data are
in reasonable agreement within uncertainties.
However, the values of $\sigma^{CLAS}$ lie in a narrower range.
Therefore, one can hope that future precise experiments could make it possible 
to distinguish between $\sigma^{PDG}$ and $\sigma^{CLAS}$.

\section{Charge asymmetry in $e^+e^-\to \eta \pi^+ \pi^-$ process}
\label{sec:asymmetry}

Though the cross section of the $e^+e^-\to f_1 \to \eta \pi^0 \pi^0$ process
is twice less than that of $e^+e^-\to f_1 \to \eta \pi^+ \pi^-$,
the former is more convenient for the study of direct $f_1$ production in
$e^+e^-$ collisions.
Indeed, the $e^+e^- \to \eta \pi^0 \pi^0$ reaction proceeds only through
two-photon annihilation, since $C$ parity of the $\eta \pi^0 \pi^0$ final
state is positive.
Therefore, there is no background from one-photon annihilation,
and the $e^+e^-\to f_1 \to \eta \pi^0 \pi^0$ cross section can be measured
directly. According to the estimate (\ref{pi0pi0}), the lower bound on this
cross section is quite small, but it can be measured in a special experiment
at the VEPP-2000 collider in Novosibirsk.

\begin{figure}[h]
\center
\begin{tabular}{c c c}
\includegraphics[scale=1.0]{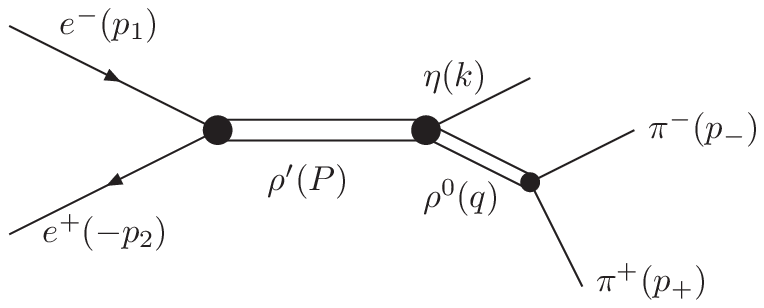} & &
\includegraphics[scale=1.0]{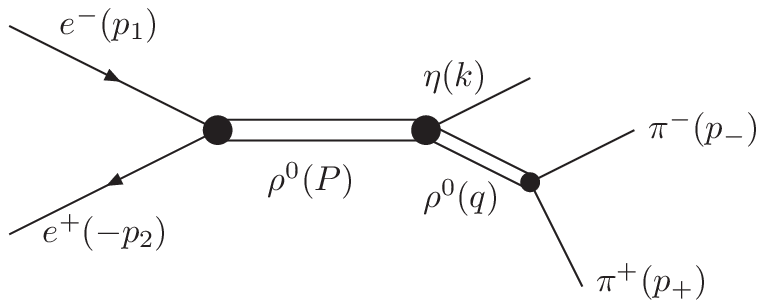}
\end{tabular}
\caption {The diagrams for $e^+e^-$ annihilation into the $\eta \pi^+ \pi^-$ final state via
intermediate vector $\rho'(1450)$ and $\rho^0 (770)$ mesons.} \label{fig:eerho}
\end{figure}

In contrast, the $e^+e^- \to \eta \pi^+ \pi^-$ reaction proceeds mainly
through one-photon annihilation, which is described quite well by the VMD
model with intermediate $\rho'(1450)$ and
$\rho^0(770)$ mesons~\cite{Aulchenko:15},
as depicted in Fig.~\ref{fig:eerho}.

The measured $e^+e^- \to \eta \pi^+ \pi^-$ Born cross section is about 500 pb
at $\sqrt{s}=m_f$~\cite{Aulchenko:15}. According to the estimate~(\ref{pi+pi-}),
the $e^+e^-\to f_1\to a_0^\pm \pi^\mp \to \eta \pi^+ \pi^-$ cross section
constitutes only several percent of the total $e^+e^- \to \eta \pi^+ \pi^-$
cross section, and its measurement is a rather complicated task.
One possibility to overcome this difficulty is to investigate the two-photon annihilation channel
$e^+e^-\to f_1 \to \eta \pi^+ \pi^-$ through $C$-odd effects, which arise
from the interference of $C$-odd one-photon and $C$-even two-photon amplitudes.

The annihilation $e^+e^- \to \rho \to \eta \pi^+ \pi^-$ was studied
theoretically in Ref.~\cite{Achasov:84}.
The corresponding formulas can also be found in Ref.~\cite{Aulchenko:15}.
The one-photon amplitude depicted in Fig.~\ref{fig:eerho} is written as
\begin{multline} \label{M1}
M_1(e^+e^- \to \eta \pi^+ \pi^-) = \frac{i f_{\rho\pi\pi}}{q^2-m_\rho^2+i\sqrt{q^2}\Gamma_\rho(q^2)}
\left( \frac{f_{\rho' ee} f_{\rho'\rho\eta}}{s-m_{\rho'}^2+i \sqrt{s}\Gamma_{\rho'}(s)} + \frac{f_{\rho ee} f_{\rho\rho\eta}}{s-m_\rho^2+i \sqrt{s}\Gamma_\rho(s)} \right) \times \\
\times \epsilon_{\lambda\nu\sigma\tau} q^\lambda p_+^\nu k^\sigma \bar{v}\gamma^\tau u.
\end{multline}
Here we take into account the dependence of $\Gamma_V$ on momentum squared,
\begin{equation}
\Gamma_{V}(s) = \Gamma_{V}(m_V^2)\frac{m_V^2}{s}\left(\frac{p_\pi(s)}{p_\pi(m_V^2)}\right)^3,
\end{equation}
where $V$ means $\rho'(1450)$ or $\rho^0 (770)$ mesons,
and $p_\pi(s) = \sqrt{s/4-m_\pi^2}$.

The coupling constant $f_{\rho\pi\pi}$ was already discussed above;
see (\ref{frhopipi}). The product $f_{Vee} f_{V\rho\eta}$
is parametrized according to Ref.~\cite{Aulchenko:15} as
$f_{Vee} f_{V\rho\eta}=4\pi\alpha m_V^2/s \cdot g_V e^{i \phi_V}$,
where $g_{\rho^0 (770)} \approx 1.58$ GeV$^{-1}$, $\phi_{\rho^0 (770)}=0$,
$g_{\rho'(1450)} \approx 0.48$ GeV$^{-1}$, and
$\phi_{\rho'(1450)}=\pi$ are obtained in Ref.~\cite{Aulchenko:15}.

The differential $e^+e^- \to \eta \pi^+ \pi^-$ cross section is written as
\begin{equation}
\frac{d\sigma(e^+e^- \to \eta \pi^+ \pi^-)}{dq^2 d\Omega_\pi d\Omega_\eta}=
\frac{1}{(2\pi)^5} \frac{1}{2s} \frac{p_\pi(q^2)}{4\sqrt{q^2}}\frac{p_\eta(q^2, s)}{4\sqrt{s}} |M(e^+e^- \to \eta \pi^+ \pi^-)|^2,
\end{equation}
where $q$ is the momentum of the $\pi^+\pi^-$ system, $\Omega_\pi$ is the solid
angle of $\pi^+$ three-momentum $\vec{p}_\pi$ in the $\pi^+\pi^-$
rest frame,
$\Omega_\eta$ is the solid angle of $\eta$ meson three-momentum
$\vec{p}_\eta$ in the $e^+e^-$ center-of-mass frame,
$p_\pi(q^2) = \sqrt{q^2/4-m_\pi^2}$, and $p_\eta(q^2, s) = \sqrt{(s-q^2-m_\eta^2)^2-4m_\eta^2 q^2}/2\sqrt{s}$.

Straightforward calculation with the amplitude (\ref{M1}) leads to the
well-known analytical formulas~\cite{Achasov:84, Aulchenko:15}.
Substituting the PDG values $m_\eta \simeq 547.862$ MeV,
$m_{\rho'(1450)} \simeq 1465$ MeV, $\Gamma_{\rho'(1450)} \simeq 400$ MeV,
and the other ones mentioned above, we obtain the following numerical result
for the cross section of the one-photon annihilation
at center-of-mass energy $\sqrt{s}=m_f=1282$ MeV  
or $\sqrt{s}=m_f^{CLAS}=1281$ MeV:
\begin{equation} \label{sigma1}
\sigma_1 (e^+e^- \to \eta \pi^+ \pi^-) \simeq 360\ {\rm pb},
\hspace{10mm} \sigma^{CLAS}_1 (e^+e^- \to \eta \pi^+ \pi^-) \simeq 350\ {\rm pb}.
\end{equation}

Now let us consider two-photon annihilation $e^+e^- \to \eta \pi^+ \pi^-$,
which proceeds (approximately with 70\% probability) via
the diagrams in Fig.~\ref{fig:eefa}.
The corresponding amplitude is as follows:
\begin{equation} \label{M2}
M_2(e^+e^- \to \eta \pi^+ \pi^-) = \frac{-i F_A \alpha^2 g_{f\pi a} g_{a\pi \eta} m_a}{s-m_f^2+i m_f \Gamma_f}\
\bar{v} \left(\frac{\hat{p}_+}{(k+p_-)^2-m_a^2 + i m_a \Gamma_a} + \frac{\hat{p}_-}{(k+p_+)^2-m_a^2 + i m_a \Gamma_a} \right) \gamma^5 u.
\end{equation}

The absolute values of the coupling constants $g_{f\pi a}$ and
$g_{a\pi \eta}$ can be found from the data on
the corresponding partial widths.
The expressions for these widths are the following:
\begin{equation}
\Gamma(f_1 \to a_0^- \pi^+) = |g_{f\pi a}|^2 \frac{\left((m_f^2-m_a^2-m_\pi^2)^2-4m_a^2 m_\pi^2\right)^{3/2}}{192\pi m_f^5},
\end{equation}
\begin{equation}
\Gamma(a_0^- \to \pi^- \eta) =  |g_{a\pi \eta}|^2 \frac{\sqrt{(m_a^2-m_\eta^2-m_\pi^2)^2-4m_\eta^2 m_\pi^2}}{16\pi m_a}.
\end{equation}
Using the experimental values 
$\Gamma(f_1 \to a_0^- \pi^+)\approx 12\% \cdot \Gamma_f \approx 2.9$ MeV, 
$m_a=980$ MeV, $\Gamma(a_0^- \to \pi^- \eta) \approx \Gamma_a \approx 60$ MeV, 
we obtain that $|g_{f\pi a}|\approx 5.23$ 
or $|g_{f\pi a}|^{CLAS}\approx 4.59$, and $|g_{a\pi \eta}|\approx 2.18$.

Since some quantities in (\ref{M2}) have large experimental uncertainties,
and the coupling constant $F_A$ depends on the free parameter
$\delta$, the value of the two-photon annihilation cross section is
quite uncertain.
Careful estimation of these uncertainties is beyond our purpose.
So, we quote here only the characteristic values of this cross section
calculated for the central values of all quantities
and for the most probable values of phase $\delta$, $\delta = 0.7\pi$, $\delta = 1.3\pi$,
and $\delta^{CLAS}=0$
(see Fig.~\ref{fig:Bf2pi2pi}),
\begin{equation} \label{sigma2}
\sigma_2 (e^+e^- \to \eta \pi^+ \pi^-) \simeq
\begin{cases}
10\ {\rm pb} & \mbox{for}\ \delta = 0.7\pi, \\
13\ {\rm pb} & \mbox{for}\ \delta = 1.3\pi,
\end{cases}
\hspace{10mm} \sigma^{CLAS}_2 (e^+e^- \to \eta \pi^+ \pi^-) \simeq 7.4\ {\rm pb}.
\end{equation}
This result is in agreement with our previous estimate (\ref{pi+pi-}).

Interference between one-photon (\ref{M1}) and two-photon (\ref{M2})
amplitudes is $P$- and $C$-odd,
therefore it does not contribute to the total cross section,
but it can lead to the {\it charge asymmetry} in the differential cross
section.
Indeed, calculation shows that after integration over azimuthal angle
$\phi_\pi$
the interference term is an odd function of $\cos{\theta_\eta}$ and
$\cos{\theta_\pi}$.
Here $\theta_\eta$ is the angle between $\eta$ meson 3-momentum
and $e^+$ beam axis in the $e^+e^-$ center-of-mass frame,
and $\theta_\pi$ is the angle between $\pi^+$ meson and $\eta$ meson 3-momenta
in the $\pi^+\pi^-$ center-of-mass system.
Therefore, if we consider events with $\theta_\eta$ in a definite interval
$d\cos{\theta_\eta}$,
then the interference term has opposite signs for $\cos{\theta_\pi}$
and $\cos{(\pi-\theta_\pi)}=-\cos{\theta_\pi}$.
Physically it means that the number of $\pi^+$ mesons propagating in
some direction $\theta_\pi$
differs from the number of $\pi^-$ mesons propagating in the same direction.

Let us define the charge asymmetry in the $e^+e^- \to \eta \pi^+ \pi^-$ process as
\begin{equation}
A=\left.\frac{\sigma_{tot}(\cos{\theta_\pi}>0) - \sigma_{tot}(\cos{\theta_\pi}<0)}{\sigma_{tot}(\cos{\theta_\pi}>0) + \sigma_{tot}(\cos{\theta_\pi}<0)}\right|_{\cos{\theta_\eta}>0},
\end{equation}
where $\sigma_{tot}=\sigma_1+\sigma_2+\sigma_{int}$ is the total cross section.
Condition $\cos{\theta_\eta}>0$ is chosen here quite arbitrarily,
so for real experiment one can redefine asymmetry in another $\theta_\eta$
range based on experimental conditions.

Since both amplitudes (\ref{M1}) and (\ref{M2}) squared are even functions
of $\cos{\theta_\eta}$ and $\cos{\theta_\pi}$,
and the interference term is an odd one, the expression for the charge asymmetry $A$ is simplified as
\begin{equation} \label{A}
A=\left.\frac{2\sigma_{int}(\cos{\theta_\pi}>0)}{\sigma_1 + \sigma_2}\right|_{\cos{\theta_\eta}>0},
\end{equation}
where the denominator is already calculated. It is one half of the sum
of (\ref{sigma1}) and (\ref{sigma2}).

\begin{figure}[h]
\center
\begin{tabular}{c c c}
\includegraphics[width=0.49\textwidth]{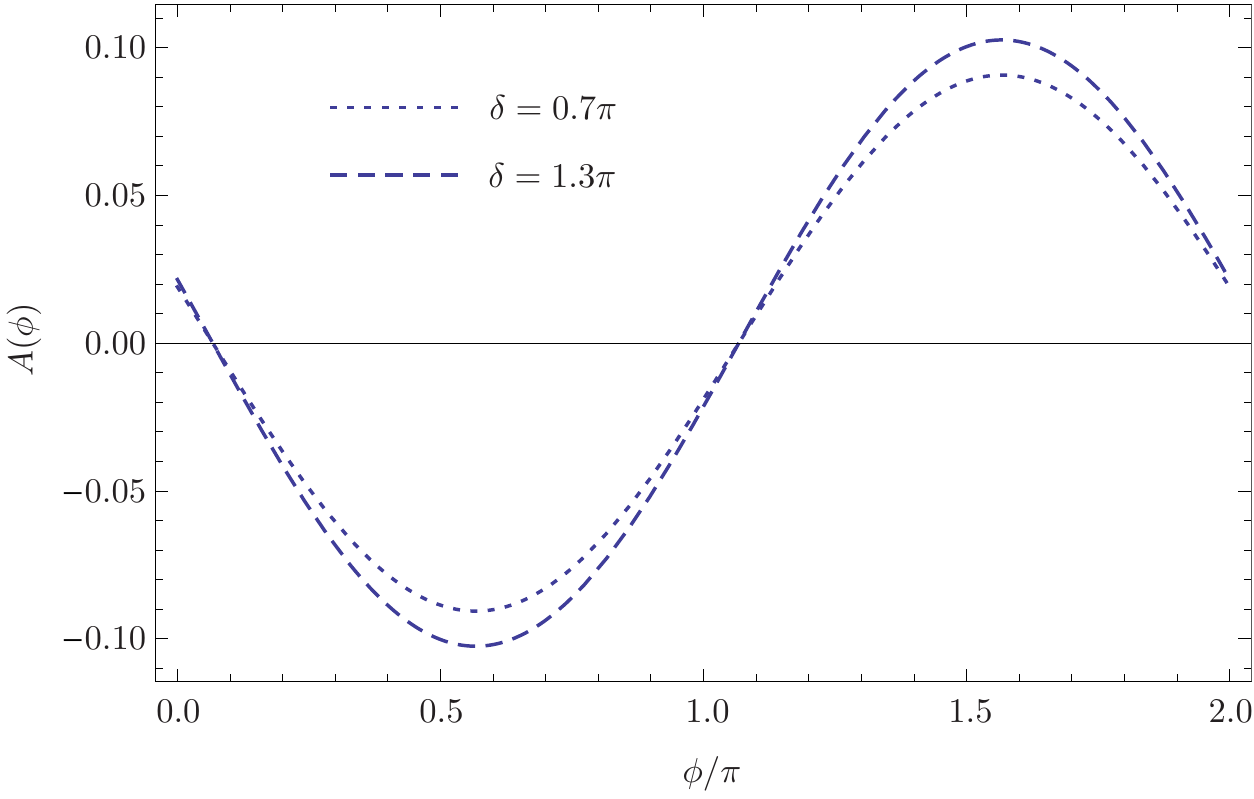} & &
\includegraphics[width=0.49\textwidth]{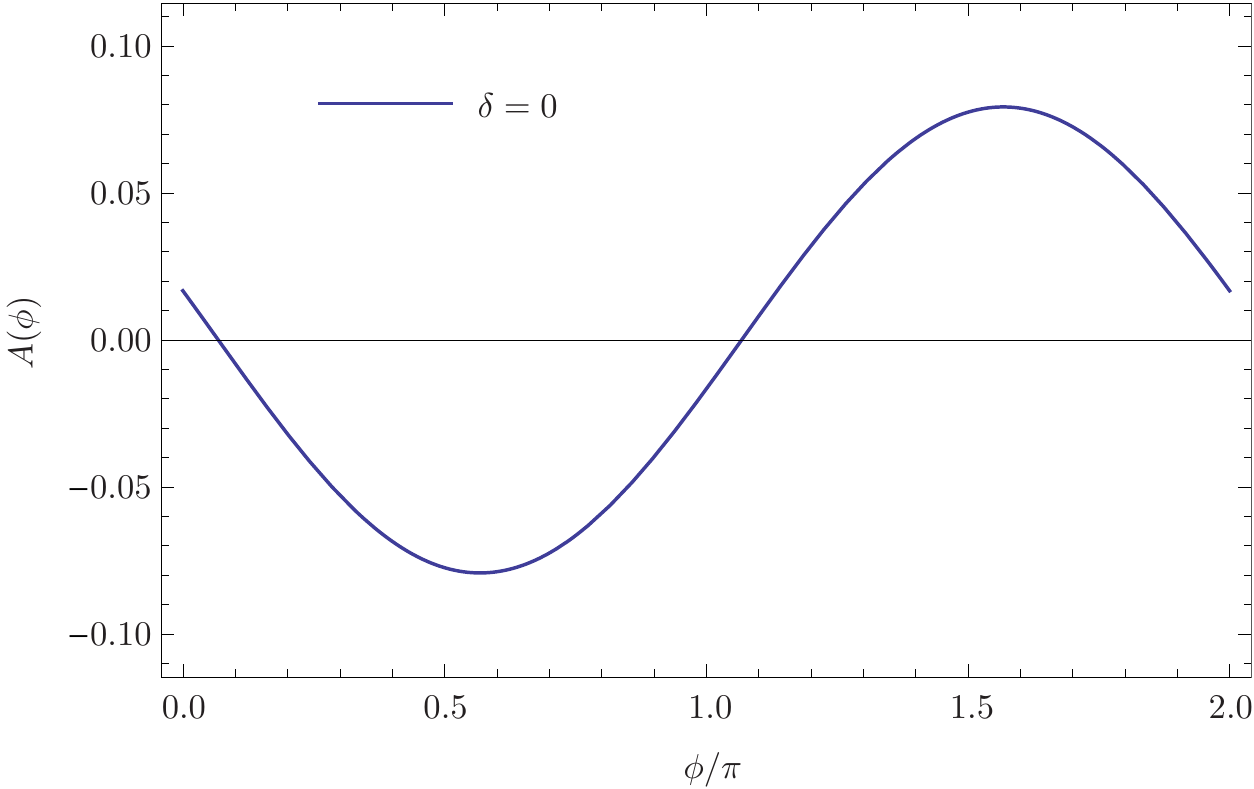}\\
PDG data & & CLAS Collaboration data
\end{tabular}
\caption {The charge asymmetry $A$ as a function of the relative phase $\phi$
for different values of the phase $\delta$.}
\label{fig:asym}
\end{figure}

The interference term contains one additional free parameter $\phi$, which
is the relative phase arising from the complex coupling constants,
\begin{equation}
F_A g_{f\pi a} g_{a\pi \eta} f_{\rho\pi\pi}^*=|F_A g_{f\pi a} g_{a\pi \eta} f_{\rho\pi\pi}|e^{i\phi}.
\end{equation}
Using the values $\phi_{\rho^0 (770)}=0$, $\phi_{\rho'(1450)}=\pi$~\cite{Aulchenko:15}
we perform numerical calculations of the charge asymmetry (\ref{A}) for 
$\delta = 0.7\pi$, $\delta = 1.3\pi$, and $\delta^{CLAS}=0$.
The dependence of the charge asymmetry $A$ on the relative phase $\phi$ is shown in Fig.~\ref{fig:asym}.

It is seen that the charge asymmetry in the $e^+e^- \to \eta \pi^+ \pi^-$ process 
may be quite large, up to $\pm 10\%$ for $\phi \simeq \mp \pi/2$.

\section{Conclusion} \label{sec:conclusion}

We calculate the width of the $f_1 (1285) \to e^+e^-$ decay in the vector meson dominance model, 
where both virtual photons are coupled with the $f_1$ meson via the intermediate $\rho^0$ mesons;
see Fig.~\ref{fig:3}.
We assume that this is the main mechanism of the decay, and such an assumption is based on the experimental data
on $f_1 \to 4\pi$ and $f_1 \to \rho^0\gamma$ decays~\cite{Barberis:00, pdg16}.
In our model the decay width, $\Gamma (f_1 \to e^+e^-)$, depends on the relative phase $\delta$ between two coupling constants describing the $f_1 \to \rho^0\gamma$ decay.
This phase is not fixed unambiguously from the experimental data.
Therefore, the width can only be estimated as $\Gamma (f_1 \to e^+e^-)\simeq 0.07 \textendash 0.19$ eV
using the PDG data~\cite{pdg16}, and as $\Gamma^{CLAS} (f_1 \to e^+e^-)\simeq 0.07 \textendash 0.10$ eV
using the CLAS Collaboration data~\cite{Dickson:16}.
The corresponding branching ratio is $\mathcal{B}(f_1 \to e^+e^-) \simeq (3 \textendash 8) \cdot 10^{-9}$
and $\mathcal{B}^{CLAS}(f_1 \to e^+e^-) \simeq (4 \textendash 5) \cdot 10^{-9}$.

The process of direct $f_1$ production in $e^+e^-$ collisions, $e^+e^-\to f_1 \to mesons$,
is still not measured due to smallness of the corresponding cross section.
Now it can be studied at modern high-luminosity colliders, e.g., at VEPP-2000 in Novosibirsk.
We estimate the $e^+e^-\to f_1 \to \eta \pi \pi$ cross section
and find it to be $\sigma(e^+e^-\to f_1\to a_0^\pm \pi^\mp \to \eta \pi^+ \pi^-)\simeq 5.2 \textendash 20\ {\rm pb}$
($\sigma^{CLAS}\simeq 7 \textendash 13.3\ {\rm pb}$) for the $\eta \pi^+ \pi^-$ final state,
and $\sigma(e^+e^-\to f_1\to a_0^0 \pi^0 \to \eta \pi^0 \pi^0)\simeq 2.6 \textendash 10\ {\rm pb}$
($\sigma^{CLAS}\simeq 3.5 \textendash 6.7\ {\rm pb}$) for the $\eta \pi^0 \pi^0$ final state.
The latter process, $e^+e^- \to f_1 \to \eta \pi^0 \pi^0$, is more convenient to study,
because the $e^+e^- \to \eta \pi^0 \pi^0$ reaction proceeds only through two-photon annihilation. 
Therefore, there is no background from one-photon annihilation,
and the $e^+e^-\to f_1 \to \eta \pi^0 \pi^0$ cross section can be measured
directly. In our model the lower bound on this cross section is quite small, $\sim 3$ pb.
However, even such a small cross section can be measured in a special experiment 
at the VEPP-2000 $e^+e^-$ collider in Novosibirsk.

In contrast, the reaction $e^+e^- \to \eta \pi^+ \pi^-$ proceeds mainly
through one-photon annihilation. 
Therefore, measurement of the cross section of the two-photon channel, 
$e^+e^-\to f_1\to \eta \pi^+ \pi^-$, 
is a rather complicated task, because of the background from one-photon annihilation.
One possibility to overcome this difficulty is to investigate 
the charge asymmetry which arises from the interference 
of $C$-odd one-photon and $C$-even two-photon amplitudes.
We calculate this asymmetry in the $e^+e^- \to \eta \pi^+ \pi^-$ reaction for 
some values of parameters in our model. 
It turns out that the magnitude of the charge asymmetry is quite uncertain.
It depends on the relative phase~$\phi$ and may be quite large, up to $\pm 10\%$.

We hope that in the nearest future our predictions will be tested in precise experiments at $e^+e^-$ colliders. 
Such experiments could allow us to obtain values of free parameters of our model, $\delta$ and $\phi$,
as well as to define more accurately $\mathcal{B}(f_1 \to \rho^0\gamma)$, $\Gamma_f$, and $r$, 
measured by now with quite large uncertainties.

\subsection*{Acknowledgments}
I am grateful to V.P.\,Druzhinin and A.I.\,Milstein for the constant interest and numerous valuable remarks and suggestions.
I also thank A.L.\,Feldman, L.V.\,Kardapoltsev, M.G.\,Kozlov, and D.V.\,Matvienko
for the useful discussions.
This work is partly supported by the Grant of President of Russian Federation
for the leading scientific Schools of Russian Federation, NSh-9022-2016.2.


\begin{thebibliography}{99}

\bibitem{Altarelli:67}
G.\,Altarelli, S.\,De Gennaro, E.\,Celeghini, G.\,Longhi, and R.\,Gatto,
{\it Theoretical calculations for electron-positron colliding-beam reactions},
Nuovo Cim. {\bf A\,47} (1967) 113.

\bibitem{Vainshtein:71}
A.I.\,Vainshtein and I.B.\,Khriplovich,
{\it On the possibility of studying resonances with positive charge parity in colliding electron-positron beams} (in Russian),
Yad. Fiz. {\bf 13} (1971) 620.

\bibitem{snd1}
M.N.\,Achasov {\it et al}. (SND Collaboration),
{\it Search for the $\eta^{\prime} \to e^+e^-$ decay with the SND detector},
Phys. Rev. {\bf D\,91} (2015) 092010 [arXiv:1504.01245].

\bibitem{kmd-3}
R.R.\,Akhmetshin {\it et al}. (CMD-3 Collaboration),
{\it Search for the process $e^+e^- \to \eta^\prime(958)$ with the CMD-3 detector},
Phys. Lett. {\bf B\,740} (2015) 273 [arXiv:1409.1664].

\bibitem{snd2}
M.N.\,Achasov {\it et al}. (SND Collaboration),
{\it Search for direct production of $a_2 (1320)$ and $f_2 (1270)$ mesons in $e^+ e^-$ annihilation},
Phys. Lett. {\bf B\,492} (2000) 8 [hep-ex/0009048].

\bibitem{besIII}
 M.\,Ablikim {\it et al}. (BESIII Collaboration),
{\it An improved limit for $\Gamma_{ee}$ of $X(3872)$ and $\Gamma_{ee}$ measurement of $\psi(3686)$},
Phys. Lett. {\bf B\,749} (2015) 414 [arXiv:1505.02559].

\bibitem{Kaplan:78}
J.\,Kaplan and J.H.\,K{\"u}hn,
{\it Direct production of $1^{++}$ states in $e^+e^-$ annihilation},
Phys. Lett. {\bf B\,78} (1978) 252.

\bibitem{Kuhn:79}
J.H.\,K{\"u}hn, J.\,Kaplan, and E.G.O.\,Safiani,
{\it Electromagnetic annihilation of $e^+e^-$ into quarkonium states with even charge conjugation},
Nucl. Phys. {\bf B\,157} (1979) 125.

\bibitem{Denig:14}
A.\,Denig, F-K.\,Guo, C.\,Hanhart, and A.V.\,Nefediev,
{\it Direct $X(3872)$ production in $e^+e^-$ collisions},
Phys. Lett. {\bf B\,736} (2014) 221 [arXiv:1405.3404].

\bibitem{Czyz:16}
H.\,Czyz, J.H.\,K{\"u}hn, and S.\,Tracz,
{\it $\chi_{c1}$ and $\chi_{c2}$ production at $e^+e^-$ colliders},
Phys. Rev. {\bf D\,94} (2016) 034033 [arXiv:1605.06803].

\bibitem{Czyz:17}
H.\,Czyz and P.\,Kisza,
{\it Testing $\chi_c$ properties at BELLE II},
Phys. Lett. {\bf B\,771} (2017) 487 [arXiv:1612.07509].

\bibitem{Kivel:16}
N.\,Kivel and M.\,Vanderhaeghen,
{\it $\chi_{cJ} \to e^+e^-$ decays revisited},
JHEP {\bf 1602} (2016) 032 [arXiv:1509.07375].

\bibitem{Kopp:74}
G.\,K{\"o}pp, T.F.\,Walsh, and P.\,Zerwas,
{\it Hadron production in virtual photon-photon annihilation},
Nucl. Phys. {\bf B\,70} (1974) 461.

\bibitem{Renard:84}
F.M.\,Renard,
{\it $1^{\pm +}$ resonances in $\gamma\gamma$ collisions},
Nuovo Cim. {\bf A\,80} (1984) 1.

\bibitem{Cahn:87}
R.N.\,Cahn,
{\it Production of spin 1 resonances in $\gamma\gamma$ collisions},
Phys. Rev. {\bf D\,35} (1987) 3342;\\
R.N.\,Cahn,
{\it Cross-sections for single tagged two photon production of resonances},
Phys. Rev. {\bf D\,37} (1988) 833.

\bibitem{Schuler:98}
G.A.\,Schuler, F.A.\,Berends, and R.\,van Gulik,
{\it Meson photon transition form-factors and resonance cross-sections in $e^+e^-$ collisions},
Nucl. Phys. {\bf B\,523} (1998) 423 [hep-ph/9710462].

\bibitem{Gidal:87}
G.\,Gidal {\it et al}. (Mark II Collaboration),
{\it Observation of spin-1 $f_1 (1285)$ in the reaction $\gamma\gamma^* \to \eta^0 \pi^+ \pi^-$},
Phys. Rev. Lett. {\bf 59} (1987) 2012.

\bibitem{Aihara:88}
H.\,Aihara {\it et al}. (TPC/2$\gamma$ Collaboration),
{\it $f_1 (1285)$ formation in photon photon fusion reactions},
Phys. Lett. {\bf B\,209} (1988) 107;\\
H.\,Aihara {\it et al}. (TPC/2$\gamma$ Collaboration),
{\it Formation of spin one mesons by photon-photon fusion},
Phys. Rev. {\bf D\,38} (1988) 1.

\bibitem{Achard:02}
P.\,Achard {\it et al}. (L3 Collaboration),
{\it $f_1 (1285)$ formation in two photon collisions at LEP},
Phys. Lett. {\bf B\,526} (2002) 269 [hep-ex/0110073].

\bibitem{Yang:12}
D.\,Yang and S.\,Zhao,
{\it $\chi_{QJ} \to l^+ l^-$ within and beyond the Standard Model},
Eur. Phys. J. {\bf C\,72} (2012) 1996 [arXiv:1203.3389].

\bibitem{pdg16}
C.\,Patrignani {\it et al}. (Particle Data Group),
{\it Review of particle physics}, Chin. Phys. {\bf C\,40} (2016) no.\,10, 100001.

\bibitem{Landau:48}
L.D.\,Landau,
{\it On the angular momentum of a system of two photons},
Dokl. Akad. Nauk USSR Ser. Fiz. {\bf 60} (1948) 207
[Collected Papers of L.D.\,Landau (Elsevier, Amsterdam, 1965), p. 471];\\
C.N.\,Yang,
{\it Selection rules for the dematerialization of a particle into two photons},
Phys. Rev. {\bf 77} (1950) 242.

\bibitem{Barberis:00}
D.\,Barberis {\it et al}. (WA102 Collaboration),
{\it A spin analysis of the $4\pi$ channels produced in central pp interactions at 450 GeV/c},
Phys. Lett. {\bf B\,471} (2000) 440 [hep-ex/9912005].

\bibitem{Amelin:95}
D.V.\,Amelin {\it et al}. (VES Collaboration),
{\it Study of the decay $f_1 (1285) \to \rho^0 (770) \gamma $},
Z. Phys. {\bf C\,66} (1995) 71.

\bibitem{Dickson:16}
R.\,Dickson {\it et al}. (CLAS Collaboration),
{\it Photoproduction of the $f_1(1285)$ meson},
Phys. Rev. {\bf C\,93} (2016) 065202 [arXiv:1604.07425].

\bibitem{Kochelev:09}
N.I.\,Kochelev, M.\,Battaglieri, and R.\,De\,Vita,
{\it Exclusive photoproduction of $f_1(1285)$ meson off the proton in kinematics 
available at the Jefferson Laboratory experimental facilities},
Phys. Rev. {\bf C\,80} (2009) 025201 [arXiv:0903.5369].

\bibitem{Wang:17}
Y.Y.\,Wang, L.J.\,Liu, E.\,Wang, and D.M.\,Li,
{\it Study on the reaction of $\gamma p \to f_1(1285) p$ in Regge-effective Lagrangian approach},
Phys. Rev. {\bf D\,95} (2017) 096015 [arXiv:1701.06007].

\bibitem{Wang:2017}
X.Y.\,Wang and J.\,He,
{\it Analysis of recent CLAS data on $f_1(1285)$ photoproduction},
Phys. Rev. {\bf D\,95} (2017) 094005 [arXiv:1702.06848].

\bibitem{Osipov:17}
A.A.\,Osipov, A.A.\,Pivovarov, and M.K.\,Volkov,
{\it Anomalous decay $f_1(1285) \to \rho \gamma$ and related processes},
Phys. Rev. {\bf D\,96} (2017) 054012 [arXiv:1705.05711].

\bibitem{Lutz:08}
M.F.M.\,Lutz and S.\,Leupold,
{\it On the radiative decays of light vector and axial-vector mesons},
Nucl. Phys. {\bf A\,813} (2008) 96 [arXiv:0801.3821].

\bibitem{Aulchenko:15}
V.M.\,Aulchenko {\it et al}. (SND Collaboration),
{\it Measurement of the $e^+e^- \to \eta \pi^+ \pi^-$ cross section in the center-of-mass energy range 1.22-2.00 GeV with the SND detector at the VEPP-2000 collider},
Phys. Rev. {\bf D\,91} (2015) 052013 [arXiv:1412.1971].

\bibitem{Achasov:84}
N.N.\,Achasov and V.A.\,Karnakov,
{\it On the research of the $e^+e^- \to \eta \pi^+ \pi^-$ reaction},
Pis'ma Zh. Eksp. Teor. Fiz. {\bf 39} (1984) 285 [JETP Lett. {\bf 39} (1984) 342].

\end{thebibliography}
\end{document}